\documentclass[twocolumn,showpacs,preprintnumbers,amsmath,amssymb]{revtex4}
\usepackage{tabularx,graphicx}\begin{document}
\newcommand{\beq}{\begin{equation}}
\newcommand{\eeq}{\end{equation}}
\newcommand{\beqn}{\begin{eqnarray}}
\newcommand{\eeqn}{\end{eqnarray}}
\newcommand{\bmath}{\begin{subequations}}
\newcommand{\emath}{\end{subequations}}
\newcommand{\rb}{\bar{r}}
\newcommand{\bk}{\bold{k}}
\newcommand{\bkp}{\bold{k'}}
\newcommand{\bq}{\bold{q}}
\newcommand{\bkb}{\bold{\bar{k}}}
\newcommand{\br}{\bold{r}}
\newcommand{\brp}{\bold{r'}}
\newcommand{\vp}{\varphi}

\title{Dynamic Hubbard model for solids with hydrogen-like atoms}
\author{J. E. Hirsch }
\address{Department of Physics, University of California, San Diego\\
La Jolla, CA 92093-0319}

\begin{abstract} 
We discuss how to construct a tight binding model Hamiltonan for the simplest possible solid, composed of hydrogen-like atoms. A single orbital per atom
is not sufficient because the on-site electron-electron repulsion mixes in higher energy orbitals. The essential physics is captured by a dynamic Hubbard model with 
one electronic orbital and an auxiliary spin degree of freedom per site. We point out that this physics can lead to a substantial shift in the position and width 
of electronic energy bands relative to what is predicted by
conventional band structure calculations.

 \end{abstract}
\maketitle 
\section{introduction}

Just like understanding the physics of the simplest atom, hydrogen, proved essential to the understanding of more complex atoms, we argue that understanding the
essential physics of a solid composed of hydrogen-like atoms is a prerequisite to understand more complex solids. When such a simplified solid is discussed it is usually
to illustrate the physics of the Mott metal-insulator transition occurring for a $1/2$-filled band\cite{mott}. Here instead we focus on the physics of energy bands with band filling larger than $1/2$, where there are necessarily some atomic orbitals / Wannier states that are doubly occupied by electrons.

Tight binding Hamiltonians commonly used to model correlated electrons in solids such as the Hubbard model, the Anderson model, the $t-J$ model and the Holstein model fail to describe the essential physical fact that {\it double occupancy of an atomic orbital changes the wavefunction of the single electron in the orbital }\cite{slater}. In fact, the wavefunction for the two electrons is not even a single Slater determinant. An approximate description with a single Slater determinant can be given, but the single electron orbital in the two-electron wavefunction is different from that in the single electron wavefunction, due to electron-electron repulsion. This is of course well known in atomic physics since at least the work of 
Hartree 85 years ago \cite{har1,har2}, but surprisingly has not yet found its way into the mainstream description of interacting electrons in solids \cite{inapp}.

In tight binding models with one orbital per site the singly occupied orbital in an atom is denoted by
\bmath
\beq
|\uparrow>=c_\uparrow^\dagger|0>
\eeq
and the doubly occupied orbital by
\beq
|\uparrow \downarrow>=c_\uparrow^\dagger c_\downarrow^\dagger |0>
\eeq
\emath
which in particular implies that
\beq
<\uparrow|c_\uparrow^\dagger|0>=<\downarrow|c_\uparrow |\uparrow \downarrow>=1
\eeq
and implies electron-hole symmetry at the atomic level, i.e. creating an electron in the empty orbital (left side of Eq. (2))  is the same as creating a hole in the doubly occupied orbital
(right side of Eq. (2)).  This is assumed to be the case in all the conventional Hamiltonians mentioned above. However
it is $qualitatively$ incorrect because the doubly occupied state is $never$ a single Slater determinant but rather a linear combination of Slater determinants involving
higher single electron states \cite{holeelec}:
\bmath
\beq
|\uparrow \downarrow>=\sum_{m,n} A_{mn}c_{m \uparrow}^\dagger c_{n\downarrow}^\dagger |0>
\eeq
\beq
\sum_{m,n}|A_{mn}|^2=1
\eeq
\emath
where the sum runs over a complete set of atomic orbitals, with the lowest single particle orbital denoted by $m=0$, i.e. $c_{0\sigma}=c_\sigma$, as well as over continuum states \cite{helium}. 
Hence
\bmath
\beq
 c_\uparrow |\uparrow \downarrow>=\sum_n A_{0n} c_{n\downarrow}^\dagger |0>=A_{00} |\downarrow> +\sum_{n\neq 0}  A_{0n} c_{n\downarrow}^\dagger |0>
\eeq
and
\beq
<\downarrow|c_\uparrow |\uparrow \downarrow> =A_{00} \neq 1
\eeq
\emath
contrary to Eq. (2). 
For the particular case $Z=2$,  the numerical value of $A_{00}$, calculated in ref. \cite{helium}, is $0.9624$. Both the fact that $A_{00}<1$ and the fact 
that there is more than one term on the right side of Eq. (4a) has important consequences for the properties of solids formed by these atoms \cite{holeelec}. Fundamentally,
the fact that Eq. (2) is invalid and Eq. (4) is valid   reflects the basic physical fact that atoms are $not$ electron-hole symmetric, nor are solids.

While this physics is ubiquitous, it is important to analyze when it will be important and when it can be neglected, just as other ubiquitous physics like  spin-orbit interaction, electron-phonon interaction, Hund's rule terms, longer-range Coulomb interactions etc are neglected in the single band Hubbard model because their effect is quantitatively small and/or because it is assumed their effects don't change the physics  qualitatively. The energy eigenvalues for an electron in the single-particle orbitals of hydrogen-like atoms are 
$\epsilon_n=-13.6( Z^2/n^2) eV$, with  $Z$ the charge of the 
nucleus. When we apply this model to more complicated atoms, $Z$ will be the $net$ charge of the
atom when there are $0$ electrons in the orbital under consideration, and $Z-1$, $Z-2$ the net
charges of the atom when there are one and two electrons in the orbital respectively.  The coefficients $A_{mn}$ in Eq. (3) can be
expressed as
\beq
A_{mn}=\frac{<\uparrow\downarrow|H_I c_{m\uparrow}^\dagger c_{n\downarrow}^\dagger |0>}
{E_0-(\epsilon_m+\epsilon_n)}
\eeq
with $H_I=e^2/|\vec{r}-\vec{r}'|$ (in first quantized form) the electron-electron interaction 
and $E_0$ the exact ground state energy of the doubly occupied orbital.
The matrix elements of $H_I$ in the numerator of Eq. (5)  are of order $Z$, while the energy differences
in the denominator are of order $Z^2$ for $(m,n)\neq(0,0)$,  
so only in the limit of large $Z$ will one have $A_{mn}\sim 0$ for $(m,n)\neq (0,0)$,  
$A_{00}\sim 1$ and $E_0\sim 2\epsilon_1+U$. As $Z$ decreases and
the electron-ion interaction becomes comparable to  the 
electron-electron interaction ($Z\sim 1$) the matrix elements $A_{mn}$ become
appreciable which implies through the sum rule Eq. (3b) that $A_{00}$ becomes
increasingly smaller. 

Put another way: using the conventional Hubbard model for the description of the
doubly-occupied  atomic states amounts to doing first order
perturbation theory in the interaction $H_I$ for the eigenvalue of the doubly occupied orbital and
zeroth order perturbation theory for the eigenstate. This is absurd unless
the spacing between non-interacting energy levels, which is of order $Z^2$, is much
larger than the matrix elements between non-interacting states and the perturbation,
which is of order $Z$ in general (provided it is not zero because of  different
angular momenta).  This is never the case in real atoms, and the effects discussed here
 will be largest for negatively charged ions
(small $Z$), where the mixing between the lowest energy states and higher energy states due to electron-electron
repulsion will be larger because of the smaller spacing between energy states and hence $A_{00}$ will be smaller.

In addition to $Z$ being small, another requirement for this physics to be important in the
solid state is that the band described by this Hamiltonian be more than half full, so that
doubly-occupied atomic states are unavoidable. For the band less than half-full,
the conventional Hubbard $U$ will keep the amplitude of doubly-occupied sites small and
this physics will not be very  important.

In recent years, `dynamic Hubbard models' have been proposed to incorporate this physics in the simplest 
possible way \cite{holeelec,dynh1,dynh2,dynh3,dynh4,dynh5,dynh6,dynh7,dynh8}. Models that have been discussed so far 
 contain either a single atomic orbital per site plus an auxiliary boson (spin 1/2 \cite{dynh2,dynh6} or harmonic oscillator \cite{dynh3,dynh5} ) degree of freedom, or two orbitals per site and no
  auxiliary boson \cite{dynh4,dynh7}. They reflect the fundamental  fact that creating an electron in the empty atomic orbital is qualitatively different
from creating a hole in the doubly occupied orbital for the reasons exposed above. 

One consequence of this physics is that   electrical conduction in an almost
full band is qualitatively different from electrical conduction in an almost empty band \cite{holeelec,dynh8}, and in particular we have shown that this can lead to pairing and superconductivity
in electronic energy bands that are almost full, driven by lowering of electronic kinetic energy \cite{holesc, kinenergy}. We have also recently shown that this leads to a tendency
of the system to expel electrons
from the interior to the surface when electronic energy bands are almost full \cite{exp1,exp2}. 

In this paper we discuss the modeling of a solid with hydrogen-like atoms with a dynamic Hubbard model with a spin degree of freedom. Extension to other
types of atoms is briefly considered.  In particular, we find  that the physics described by the dynamic Hubbard model will  lead to a shift in the position and width of energy bands 
that are full or almost full relative to what
is predicted by conventional band structure calculations.

\section{the hydrogen-like atom}

The simplest non-trivial form for the ground state wave function of two electrons in a
hydrogen-like atom is of the Hartree form
\bmath
\beq
\Psi(r_1,r_2)=\varphi_{\tilde{Z}}(r_1)\varphi_{\tilde{Z}}(r_2)
\eeq
\beq
\varphi_{\tilde{Z}}(r_1)=\sqrt{\frac{\tilde{Z}^3}{\pi}}e^{-\tilde{Z}r/a_0}
\eeq
\emath
with $a_0=\hbar^2/(m_e e^2)$ the Bohr radius. 
This is an approximation to the linear combination of Slater determinants given by Eq. (3). 
For the singly occupied orbital the wavefunction is Eq. (6b) with $\tilde{Z}=Z$,
with $Ze$ the nuclear charge.
For the doubly occupied orbital the kinetic and  potential energies for each electron and the 
electron-elsctron   interaction energy are given by
\bmath
\beq
E_{kin}(\tilde{Z})=\frac{\hbar^2}{2m_e a_0^2}\tilde{Z}^2=\frac{e^2}{2a_0}\tilde{Z}^2
\eeq
\beq
E_{pot}(\tilde{Z})=-\frac{e^2}{a_0} Z\tilde{Z}
\eeq
\beq
E_{ee}(\tilde{Z})=\frac{5}{8}\frac{e^2}{a_0}\tilde{Z}
\eeq
\emath
so that the total energy for the two-electron ion is
\beq
E_{tot}=\frac{e^2}{a_0} (\tilde{Z}^2-2Z\tilde{Z}+\frac{5}{8}\tilde{Z})
\eeq
and is minimized by
\beq
\tilde{Z}=Z-\frac{5}{16}\equiv Z-\delta \equiv \bar{Z}\
\eeq
so that the wavefunction expands upon double occupancy from radius $a_0/Z$ to radius $a_0/\bar{Z}$. The spatial extent of the wavefunction
 becomes larger the smaller the ionic charge
$Z$, and diverges as $Z\rightarrow 5/16=0.3125$.

 \begin{figure}
\resizebox{8.5cm}{!}{\includegraphics[width=7cm]{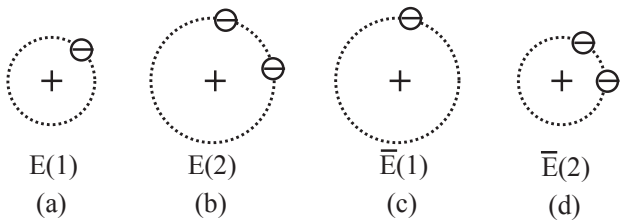}}
\caption {  States (a) and (b) are the lowest energy states for 1 and 2 electrons, with unexpanded and expanded orbital respectively. States (c) and (d) are 
 one and two-electron states with expanded and unexpanded orbital respectively, and correspondingly higher energies. In the conventional
 Hubbard model only states (a) and (d) are considered.}
\label{figure1}
\end{figure}

Figure 1 shows schematically the four states of interest here. (a) and (b) are the ground states of the atom with one and two-electrons respectively,
with unexpanded and expanded orbits respectively and energies
\bmath
\beq
E(1)=E_{kin}(Z)+E_{pot}(Z)=-\frac{e^2}{2a_0} Z^2
\eeq
\beqn
E(2)&=&2(E_{kin}(\bar{Z})+E_{pot}(\bar{Z}))+E_{ee}(\bar{Z}) \nonumber \\
&=& -\frac{e^2}{a_0} (Z^2-2\delta Z+\delta^2 )
\eeqn
\emath
The states (c) and (d) are excited states (but they are $not$ orthogonal to the ground
states): (c) for the single electron in the expanded orbit, and (d) for the two electrons in the non-expanded orbit, with energies
\bmath
\beq
\bar{E}(1)= E_{kin}(\bar{Z})+E_{pot}(\bar{Z})=-\frac{e^2}{2a_0}( Z^2-\delta^2)
\eeq
\beqn
\bar{E}(2)&=&2(E_{kin}(Z)+E_{pot}(Z))+E_{ee}(Z) \nonumber \\
&=& -\frac{e^2}{a_0} Z(Z-2\delta)
\eeqn
\emath
respectively. We note that the energies satisfy
\bmath
\beq
\bar{E}(1)-E(1)=\frac{\delta^2}{2} \frac{e^2}{a_0}
\eeq
\beq
\bar{E}(2)-E(2)=\delta^2\frac{e^2}{a_0}
\eeq
\emath
so the difference in the energy of the excited state and the ground state is independent of $Z$, and in addition they satisfy
\beq
\bar{E}(2)-E(2)=2(\bar{E}(1)-E(1)) .
\eeq
The right side of Eq. (13) is twice the cost in single electron energy in expanding the orbital from its single electron radius to the larger radius apropriate to the 
two-electron atom. The two-electron atom pays that cost but gains twice as much from the reduction in electron-electron repulsion
achieved by expanding the orbital:
\beq
E_{ee}(Z)-E_{ee}(\bar{Z})=2\delta^2 \frac{e^2}{a_0}=4(\bar{E}(1)-E(1)).
\eeq
From this point of view it can be said that the orbital expansion is driven by lowering of the electron-electron interaction energy at a cost of single-particle energy.

It is also interesting to ask whether the expansion is `kinetic energy driven' or `potential energy driven'. Upon double occupancy and resulting orbital expansion the electron-ion
potential energy increases more than the reduction in electron-electron repulsion, giving rise to a net $increase$ in potential energy:
\beqn
\bar{E}_{pot}-E_{pot}&=&2(E_{pot}(\bar{Z})-E_{pot}(Z))+E_{ee}(\bar{Z})-E_{ee}(Z) \nonumber \\
&=&  2\delta \frac{e^2}{a_0}(Z-\delta)
\eeqn
which is always positive due to the requirement that $\bar{Z}>0$. On the other hand the kinetic energy always decreases
\beq
\bar{E}_{kin}-E_{kin}=2(E_{kin}(\bar{Z})-E_{kin}(Z))=-2\delta \frac{e^2}{a_0}(Z-\frac{\delta}{2})
\eeq
for a total energy lowering
\beq
\bar{E}_{tot}-E_{tot}=E(2)-\bar{E}(2)=-\delta^2\frac{e^2}{a_0}
\eeq
so one can say that the orbital expansion upon double occupancy is always ``kinetic energy driven''.
Also, in comparing the lowering of kinetic energy Eq. (16) with the lowering of electron-electron repulsion energy Eq. (14) we find that the former is larger as long as
\beq
Z>\frac{3}{2}\delta
\eeq
or $\bar{Z}>\delta/2=0.15625$, a rather small value. The ratio of kinetic energy lowering to electron-electron energy lowering is
\beq
\frac{\Delta E_{kin}}{\Delta E_{ee}}=\frac{2Z}{\delta}-\frac{1}{2}
\eeq
so that for hydrogen for example ($Z=1$) the lowering of kinetic energy upon double occupancy and resulting orbital expansion is more than
three times larger than the lowering of electron-electron repulsion energy.

In the conventional Hubbard model the orbital expansion is not considered. The value of the Hubbard $U$ in the unrelaxed atomic orbital is
\beq
U_{bare}=E_{ee}(Z)= \frac{5}{8}\frac{e^2}{a_0}Z
\eeq
In the expanded orbital the electron-electron repulsion is lower, given by
\beq
U_{exp}=E_{ee}(\bar{Z})=\frac{5}{8} \frac{e^2}{a_0}(Z-\delta)
\eeq
However the actual ``effective $U$" is larger than $U_{exp}$, since the single particle energy cost in expanding the orbital has to be taken into account. 
The effective $U$ is given by
\beq
U_{eff}=E(2)-2E(1)=\frac{5}{8} \frac{e^2}{a_0}(Z-\frac{\delta}{2})
\eeq
so that it is the average of the repulsions in the expanded and unexpanded orbitals. Note that $U_{bare}$ and $U_{exp}$ are also given by the expressions
\bmath
\beq
U_{bare}=\bar{E}(2)-2E(1)
\eeq
\beq
U_{exp}=E(2)-2\bar{E}(1)  .
\eeq
\emath

 \begin{figure}
\resizebox{8.5cm}{!}{\includegraphics[width=7cm]{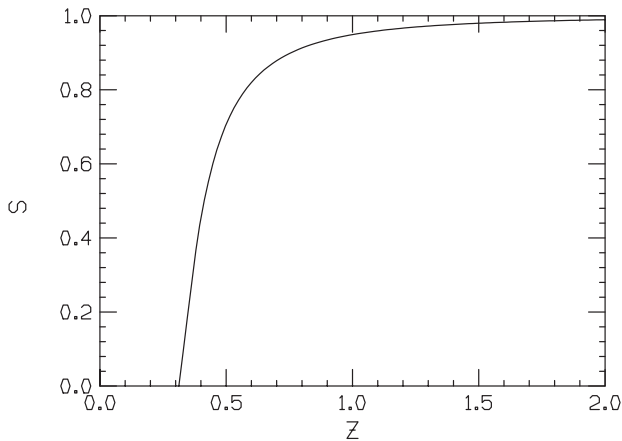}}
\caption {  Overlap matrix element of the expanded and unexpanded orbital $S$ versus atomic charge $Z$  }
\label{figure1}
\end{figure} 
Finally, it is of interest to compute the overlap matrix element of the single particle expanded and unexpanded orbital
\beq
S=<\varphi_Z|\varphi_{\bar{Z}}>=
\frac{(Z \bar{Z})^{3/2}}{(\frac{Z+\bar{Z}}{2})^3} =
\frac{(1-\frac{\delta}{Z})^{3/2}}{(1-
\frac{\delta}{2Z})^3} .
\eeq
This is plotted in Fig. 2 as function of the ionic charge $Z$. The overlap decreases as the ionic charge decreases and approaches zero for $Z\rightarrow 5/16$ where the expanded
orbital becomes infinitely large. Note however that the Hartree approximation $underestimates$ the magnitude of these
effects. 
For example, for $Z=2$ the Hartree wave function  gives $S=0.9892$ and the highly accurate Hylleraas wave function that includes radial and angular correlations and should be essentially exact gives $S=0.9810$.
This value is obtained from the square root of the overlap matrix element of the
two-electron wave function with the wave function of two electrons in the unexpanded orbitals.

The matrix element $S$ plays a key role in determining the properties of electrons in the energy band generated by this atomic orbital \cite{dynh5}:
the quasiparticle weight decreases and the effective mass of the carriers increases as the band filling increases, with the magnitude of these effects determined by the
deviation of $S$ from unity.

 \section{dynamic Hubbard model with auxiliary spin degree of freedom}
 
\subsection{Simplest model}

The simplest form for the site Hamiltonian for a dynamic Hubbard model with an auxiliary spin 1/2 degree of freedom is \cite{dynh6}
\bmath
\beqn
H_i=&\epsilon_e&+\epsilon_0(n_{i\uparrow}+n_{i\downarrow}) + \omega_0\sigma_x^i +g\omega_0\sigma_z^i \nonumber \\
&+&(U_0- 2g\omega_0\sigma_z^i ) n_{i\uparrow} n_{i\downarrow}
\eeqn
\beq
\epsilon_e= \omega_0\sqrt{1+g^2}
\eeq
\emath
 so that the (lowest) energy is zero when there are no electrons at the site. The spin part of the Hamiltonian is
\bmath \beq
H_{spin}(n=0, n=1)=\omega_0\sigma_x^i +g\omega_0\sigma_z^i
\eeq
\beq
H_{spin}(n=2)=\omega_0\sigma_x^i -g\omega_0\sigma_z^i
\eeq
\emath
when there are  $n$ electrons at the site. The eigenvalues of these spin Hamiltonians are
\beq
\epsilon=\pm \omega_0 \sqrt{1+g^2} .
\eeq
and the eigenvectors
\bmath
\beq
|n>=u(n)|+>+v(n)|->
 \eeq
 \beq
|\bar{n}>=v(n)|+>-u(n)|->
 \eeq
 \emath
 with $|n>$ denoting the ground state and $|\bar{n}>$ the excited state for the spin with $n$ electrons at the site, $|+>$ and $|->$ the basis states for the spin degree of 
 freedom, and
 \bmath
 \beq
 u(n)\equiv u=\sqrt{\frac{1}{2}(1-\frac{g}{\sqrt{1+g^2}})} 
 \eeq
 \beq
 v(n)\equiv v=-\sqrt{\frac{1}{2}(1+\frac{g}{\sqrt{1+g^2}})} .
 \eeq
 \emath
 for $n=0, 1$ and
 \bmath
 \beq
 u(2)=v(1)
 \eeq
 \beq
 v(2)=u(1)  .
 \eeq
 \emath
 
 This Hamiltonian is particularly simple because  the eigenvalues Eq. (27) are independent of the site occupation and the eigenvectors for different occupations are related in the simple way given by
 Eq. (30), and for this reason it was chosen in Ref. \cite{dynh6}. Later in this section we consider  other possible versions of this model where these expressions take a less
 simple form.

 The essential physics of this (and the other versions of this) Hamiltonian is that the lowest state of the spin is the same for $0$ or $1$ electron at the site, and different for $2$ electrons at the site. For large
 $g$,  the lowest energy states are approximately $|->$ for  $n=0$ and $n=1$ and $|+>$ for $n=2$. The change in the state of the spin represents the orbital expansion
 as shown schematically in Fig. 3. The overlap
 matrix elements between small and large orbitals Eq. (24) is thus given by
 \beq
 S=<1|2>=|2uv|=\frac{1}{\sqrt{1+g^2}}
 \eeq
 so that $g$ is given by
 \beq
 g=\sqrt{\frac{1}{S^2}-1}
 \eeq
 The lowest site energies in the dynamic Hubbard model Eq. (25) are
 \bmath
 \beq
 E(0)=0
 \eeq
 \beq
 E(1)=\epsilon_0 
 \eeq
 \beq
 E(2)=2\epsilon_0+U_0 
 \eeq
 \emath
 
    \begin{figure}
\resizebox{8.5cm}{!}{\includegraphics[width=7cm]{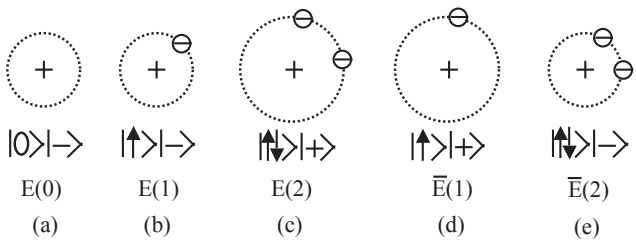}}
\caption { Dynamic Hubbard model representation of the atomic states. The $|->$ and  $|+>$spin states correspond to the unexpanded and expanded orbital
in the limit $g\rightarrow \infty$. For finite $g$, they are replaced by the states $|1>$ and  $|2>$ respectively (see text).  
}
\label{figure1}
\end{figure}

If we destroy an electron from the ground state of the two-electron atom the spin remains in state $|2>$. The expectation value of the Hamiltonian for one electron
with the spin state $|2>$ then gives  the energy of one electron in the expanded orbital, $\bar{E}(1)$:
\beq
\bar{E}(1)=<2|H(n=1)|2>=\epsilon_0+\omega_0\sqrt{1+g^2}+\frac{g^2-1}{\sqrt{1+g^2}}  \omega_0
\eeq
and similarly the expectation value of the two-electron Hamiltonian with the spin state $|1>$ gives the energy of the two-electron atom with the unexpanded orbital:
\beqn
\bar{E}(2)&=&<1|H(n=2)|1> = 2\epsilon_0+U_0 \nonumber \\
&+& \omega_0\sqrt{1+g^2} + \frac{g^2-1}{\sqrt{1+g^2}}  \omega_0
\eeqn
Note however  that this implies
\beq
\bar{E}(2)-E(2)=\bar{E}(1)-E(1)
\eeq
which disagrees with Eq. (13) obeyed by the atomic energies. Thus, it is not possible to find parameters in the model Eq. (25)  that will match the four energies of the atomic states
considered. 
So we ignore the energy
$\bar{E}(2)$ and determine the parameters in the model by equating the energies $E(1)$, $E(2)$ and $\bar{E}(1)$ to the atomic energies and by Eq. (32). The result is
\bmath
\beq
\omega_0=\frac{25S}{1024(1-S^2)}\frac{e^2}{a_0}
\eeq
\beq
 \epsilon_0=-\frac{e^2}{2a_0}Z^2
\eeq
\beq
 U_0=U_{eff}=\frac{5}{8}\frac{e^2}{a_0}(Z-\frac{5}{32})
\eeq
\emath
Thus, Eqs. (32) and (37) determine the parameters in the dynamic Hubbard model Hamiltonian as function of the atomic parameters. 

Let us consider some numerical examples. (a) For $Z=1$, $S=0.949$ and the parameters in the model are $g=0.332$, $\omega_0=6.34eV$, $\epsilon_0=-13.6eV$,
$U_0=14.35eV$. (b) For $Z=0.3557$, $S=0.5$ and the parameters in the model are $g=1.732$, $\omega_0=0.443eV$, $\epsilon_0=-1.72eV$, 
$U_0=3.39eV$. 

\subsection{Alternative form of the model}

We can  generalize the model Eq. (25) to

\beqn
H_i=&\epsilon_e&+\epsilon_0(n_{i\uparrow}+n_{i\downarrow}) + \omega_0\sigma_x^i +g\omega_0\sigma_z^i \nonumber \\
&+&(U_0-a g\omega_0\sigma_z^i) n_{i\uparrow} n_{i\downarrow} .
\eeqn
with $\epsilon_e$ is still given by Eq. (25b), so that $E(0)=0$. 
As long as the parameter $a>1$ the physics is essentially the same as that of Eq. (25)
(which is the case $a=2$), particularly for large $g$: the state of the spin changes drastically when a hole 
is created at the doubly occupied site but doesn't change when an electron is created at the empty site. The eigenvalues of the spin Hamiltonian are now
\bmath
\beq
\epsilon=\pm \omega_0\sqrt{1+g^2}
\eeq
for $0$ or $1$ electron at the site, and 
\beq
\epsilon=\pm \omega_0\sqrt{1+(a-1)^2g^2}
\eeq
\emath
for $2$ electrons at the site, so the
eigenvalues are no longer independent of site occupation for any $a\neq 2$.

Introducing the parameter $a$ allows us to satisfy the condition Eq. (13). It will be satisfied if the relation
\beq
\sqrt{1+(a-1)^2g^2}=2\sqrt{1+g^2}
\eeq
holds, hence if $a$ is given by 
\beq
a=a(g)=1+\sqrt{4+\frac{3}{g^2}} .
\eeq
$a$ increases monotonically as $g$ decreases, 
from $a=3$ for $g\rightarrow \infty$ to $a(g)\sim \sqrt{3}/g$ for  $g\rightarrow 0$ .  
The energies are given by
\bmath
\beq
E(1)=\epsilon_0 
\eeq
\beq
\bar{E}(1)=\epsilon_0+\frac{1+2g^2+g\sqrt{3+4g^2}}{2\sqrt{1+g^2}}\omega_0
\eeq
\beq
E(2)=2\epsilon_0+U_0-\omega_0\sqrt{1+g^2}
\eeq
\beq
\bar{E}(2)=2\epsilon_0+U_0+ \frac{g^2+g\sqrt{3+4g^2}}{\sqrt{1+g^2}}\omega_0
\eeq
\emath
so that the atomic relation $\bar{E}(2)-E(2))=2(\bar{E}(1)-E(1))$ holds for all $g$. 
In the limit of large $g$, $a=3$ and we have
$
\bar{E}(1)=\epsilon_0+2g\omega_0
$, $
E(2)=2\epsilon_0+U_0-g\omega_0
$, $
\bar{E}(2)=2\epsilon_0+U_0+3g\omega_0$.
For any $g$ the eigenvector amplitudes  are given by
\bmath
\beq
u(2)=\sqrt{\frac{1}{2}(1+\frac{\sqrt{3+4g^2}}{2\sqrt{1+g^2})})}
\eeq
\beq
v(2)=-\sqrt{\frac{1}{2}(1-\frac{\sqrt{3+4g^2}}{2\sqrt{1+g^2})})}
\eeq
\emath

The coupling constant $g$ is again determined from the condition $S=<1|2>$ which now is
\beq
S=\frac{\sqrt{3+2g^2-g\sqrt{3+4g^2}}}{2\sqrt{1+g^2}}
\eeq
with $S$ given by Eq. (24). It can be seen from Eq. (44) that as $g\rightarrow 0$, $S\rightarrow \sqrt{3}/2=0.866$.  In that limit  the spin part of the Hamiltonian is
\bmath
\beq
H_i^{spin}(n=1)=\omega_0\sigma_x^i
\eeq
\beq
H_i^{spin}(n=2)=\omega_0\sigma_x^i-\sqrt{3}\omega_0\sigma_z^i
\eeq
\emath
with energies $\epsilon(1)=\pm \omega_0$ and $\epsilon(2)=\pm 2\omega_0$
giving rise to $<1|2>=\sqrt{3}/2$ and satisfying the condition Eq. (13).

Therefore, we conclude that the Hamiltonian of the form Eq. (38) with $a$ chosen to satisfy the condition Eq. (13) can only be used to match the 4 atomic energies Eqs. (10) and (11) provided $S<0.866$, which from 
Eq. (24) corresponds to $Z<0.673$. In any event the regime of the model that gives rise to interesting properties in the solid is large $g$, corresponding to
small $S$ and $Z$.

So to find the appropriate parameters in the model for a given $Z<0.673$ and corresponding $S$ given by Eq. (24) we first find $g$ from Eq. (44).
$\epsilon_e$ and $\epsilon_0$ are still given by Eqs. (25b) and (37b), and the other parameters are given by
\bmath
\beq
\omega_0=\frac{25}{256} \frac{e^2}{a_0}\frac{\sqrt{1+g^2}}{1+2g^2+\sqrt{3g^2+4g^4}}
\eeq
\beq
U_0=\frac{5}{8}\frac{e^2}{a_0}(Z-\frac{5}{32})+\omega_0\sqrt{1+g^2} .
\eeq
\emath

For the  numerical example given in the previous section $Z=0.3557$, $S=0.5$, we obtain for the parameters in this model $g=2/\sqrt{3}=1.1554$,
$\omega_0=25e^2/(256\sqrt{21}a_0)=0.580 eV$, $\epsilon_0=-1.72eV$, $U_0=4.28eV$, 
versus $g=1.732$, $\omega_0=0.443eV$, $\epsilon_0=-1.72eV$, 
$U_0=3.39eV$ in the simpler model Eq. (25).

\subsection{Third version of the model}
 Finally we consider also the Hamiltonian Eq. (38) with $a=3$ independent  of  the value of $g$. Recall that this value of $a$ satisfies the atomic condition Eq. (13) only for $g\rightarrow \infty$. However, the deviation from this condition is much
smaller than for the model of the form Eq. (25). The lowest eigenvalues for different occupations are
$E(1)=\epsilon_0$  and 
\beq
E(2)=2\epsilon_0+U_0+\omega_0\sqrt{1+g^2} -\omega_0\sqrt{1+4g^2}
\eeq
and the energies for one electron in the expanded orbital and two electrons in the unexpanded orbital are respectively
\bmath
\beq
\bar{E}(1)=\epsilon_0+\omega_0\sqrt{1+g^2} +\frac{2g^2-1}{\sqrt{1+4g^2}}\omega_0
\eeq
\beq
\bar{E}(2)=2 \epsilon_0+U_0+\frac{3g^2}{\sqrt{1+g^2}}\omega_0
\eeq
\emath
The relation Eq. (13) is no longer satisfied, rather
\bmath
\beq
\bar{E}(2)-E(2)=C(\bar{E}(1)-E(1))
\eeq
\beq
C=\sqrt{\frac{1+4g^2}{1+g^2}}
\eeq
\emath
so for example $C=1.96, 1.92, 1.84, 1.58$ for $g=4, 3, 2, 1$, closer to the atomic value $C=2$ than in the model Eq. (25) for which $C=1$ for all $g$.

 The eigenvectors are given by Eq. (29) for $n=0, 1$, and for $n=2$ by
\bmath
 \beq
 u(2) =\sqrt{\frac{1}{2}(1+\frac{2g}{\sqrt{1+4g^2}})} 
 \eeq
 \beq
 v(2) =-\sqrt{\frac{1}{2}(1-\frac{2g}{\sqrt{1+4g^2}})} ,
 \eeq
 \emath
The coupling constant $g$ is determined from the condition $S=<1|2>$ which is now
\beq
S=\sqrt{\frac{1}{2}(1+\frac{1-2g^2}{\sqrt{1+5g^2+4g^4}})}
\eeq
and the other parameters in the Hamiltonian are given by Eq. (37b) and 
\bmath
\beq
\omega_0=\frac{25}{512}\frac{e^2}{a_0}\frac{\sqrt{1+4g^2}}{2g^2-1+\sqrt{(1+g^2)(1+4g^2)}}
\eeq
\beq
U_0=\frac{5}{8}\frac{e^2}{a_0}(Z-\frac{5}{32})+\omega_0\sqrt{1+4g^2}-\omega_0\sqrt{1+g^2}   .
\eeq
\emath

 \begin{figure}
\resizebox{8.5cm}{!}{\includegraphics[width=7cm]{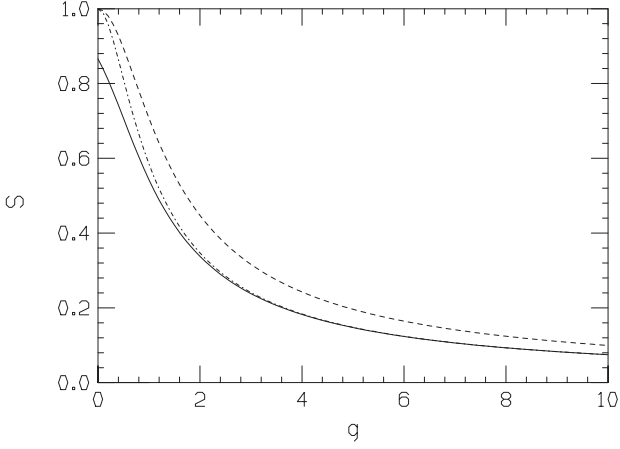}}
\caption { Overlap matrix element $S$ versus coupling constant $g$ for the first (Eq. 25, dashed line), second (Eq. 38, solid line) and third (Eq. 38 with $a=3$, dash-dotted line)  version of
the model  }
\label{figure1}
\end{figure}

 \begin{figure}
\resizebox{8.5cm}{!}{\includegraphics[width=7cm]{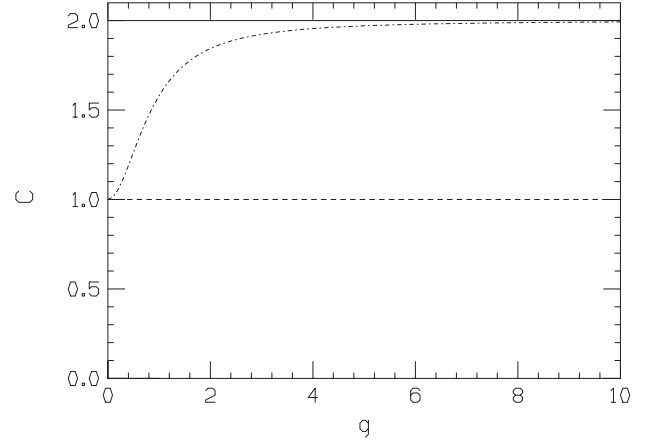}}
\caption { Ratio of energies $C=(\bar{E}(2)-E(2))/(\bar{E}(1)-E(1))$ versus coupling constant $g$ for the first (Eq. 25, dashed line), second (Eq. 38, solid line) and third (Eq. 38 with $a=3$, dash-dotted line) version of
the model. In the atom, $C=2$ for all values of the coupling constant.}
\label{figure1}
\end{figure} 

  \begin{figure}
\resizebox{8.5cm}{!}{\includegraphics[width=7cm]{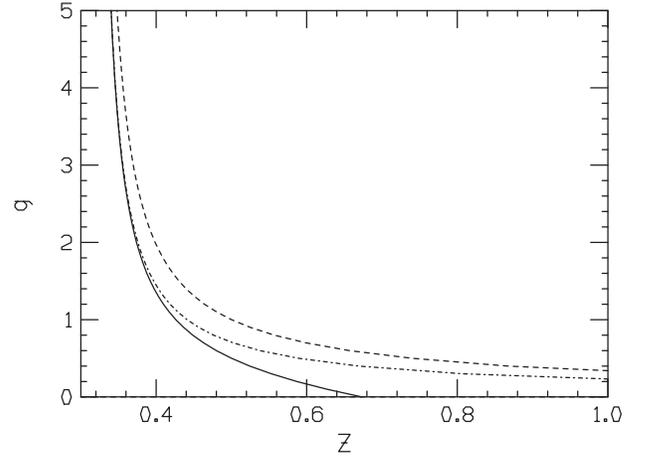}}
\caption { Coupling constant $g$ versus atomic charge $Z$ for the first (Eq. 25, dashed line), second (Eq. 38, solid line) and third (Eq. 38 with $a=3$, dash-dotted line) version of
the model  }
\label{figure1}
\end{figure} 

Figure 4  shows the overlap matrix element $S$ versus coupling constant $g$, fig. 5 shows the ratio of energy differences $C$  (Eq. (49)) and fig. 6 shows the coupling constant $g$ versus
atomic charge $Z$ for the three versions of the model. The second and third version become equivalent in the strong coupling regime which corresponds to the case of
negatively charged ions (small $Z$).

\section{effective Hamiltonian}
The Hamiltonian for the lattice system is
\beq
H=\sum_i H_i -t\sum_{i,j}    (c_{i\sigma}^\dagger c_{j\sigma}+h.c.)
\eeq
acting on the product Hilbert space of electron states and spin states. We denote the site states by the product of electron and spin states. The action of a fermion 
creation operator is
\bmath
\beq
c_{i\uparrow}^\dagger |0>|0>=|\uparrow>|0>=|\uparrow>|1>
\eeq
\beq
c_{i\uparrow}^\dagger |\downarrow>|1>=  S |\uparrow\downarrow>|2>+\bar{S}|\uparrow\downarrow>|\bar{2}>
\eeq
\emath
with
$S=<1|2>$, $\bar{S}=<1|\bar{2}>$. $|\bar{2}>$ denotes the higher energy spin state for two electrons at the site:
\beq
|\bar{2}>=v(2)|+>-u(2)|->  .
\eeq
The first term in Eq. (54b) describes the ground-state to ground-state (diagonal) transition, i.e. the process where the orbital expands when doubly occupied.
The second term is the process where the doubly occupied site ends up in an excited state. 
We define quasiparticle fermion operators $\tilde{c}_{i\sigma}^\dagger$  by the relation
\beq
c_{i\sigma}^\dagger=[1-(1-S)\tilde{n}_{i,-\sigma}]\tilde{c}_{i\sigma}^\dagger+\bar{S}|\uparrow \downarrow>|\bar{2}><1|<\sigma|
\eeq
to describe the diagonal transitions and  to obtain the effective low energy Hamiltonian we ignore the second term in Eq. (56):
\bmath
\beqn
H_{eff}&=&  
  \sum_{ij} t_{ij}^\sigma  (\tilde{c}_{i\sigma}^\dagger \tilde{c}_{j\sigma}+h.c.) \nonumber \\
&+&U_{eff}\sum_i  \tilde{n}_{i\uparrow}  \tilde{n}_{i\downarrow}+\epsilon_0\sum_i( \tilde{n}_{i\uparrow}+ \tilde{n}_{i\downarrow})
\eeqn
\beq
t_{ij}^\sigma=t[1-(1-S)\tilde{n}_{i,-\sigma}][1-(1-S)\tilde{n}_{i-\sigma}]
\eeq
\emath
to describe the low-energy physics. 

As discussed elsewhere \cite{holesc}, this Hamiltonian gives rise to high temperature superconductivity when
the band is almost full and the overlap matrix element $S$ is sufficiently small. The condition for superconductivity for an almost full band is \cite{polaronic}
\beq
S \leq \sqrt{1-\frac{U_{eff}}{D}}
\eeq
with $D$ the unrenormalized bandwidth ($D=2zt$, with $z$ the number of nearest neighbors to a site). Note that as the net ionic charge $Z$ decreases
the overlap matrix element $S$ decreases (Fig. 2) and in addition $U_{eff}$ (Eq. (22)) decreases, so that the condition Eq. (65) is more easily satisfied.
Thus according to this model  high temperature superconductivity  is favored by having negatively charged ions.

The factor $S^2$ represents the quasiparticle weight when the
band is almost full \cite{dynh5}. 
The second term in Eq. (54b) gives rise to
incoherent processes that contribute to the high energy optical absorption and photoemission spectra for bands that are almost full \cite{holeelec}.
Thus, high temperature superconductivity in this model is associated with small quasiparticle weight and large spectral weight for incoherent
processes ($S^2+\bar{S}^2=1$), physical features found  in high $T_c$ cuprates
 \cite{incoh0,incoh1,incoh2}.

\section{Shift of energy bands}

We argue that the orbital relaxation effect described by the dynamic Hubbard model will shift the location of energy bands that are almost full relative to what would be 
predicted by density functional band structure calculations.

Consider a band structure calculation within density functional theory (DFT) that predicts a full band several $eV$ below the Fermi energy $\epsilon_F$, as shown
in Fig. 7. The calculation assumes that each atomic orbital is doubly occupied, and predicts that the Fermi level has to come down a distance
$\Delta^{DFT}$ for doped holes to go into that band, hence that doped holes will go into other bands that cross the Fermi energy. The distance between the top of the band
and the Fermi energy is
\beq
\Delta^{DFT}=E_{final}^{DFT}-E_{initial}
\eeq
where $E_{initial}$ and $E_{final}^{DFT}$ are the initial and final energies of the system upon bringing an electron from the top of the band to the Fermi energy.

  \begin{figure}
\resizebox{8.5cm}{!}{\includegraphics[width=7cm]{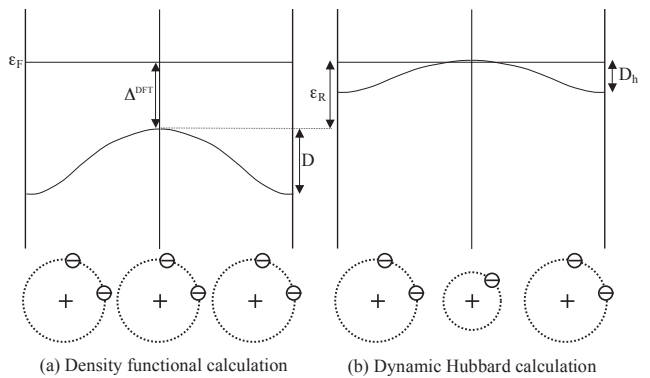}}
\caption { In a standard band structure calculation using density functional theory (a), the orbital relaxation energy when an electron is removed from the band is not
taken into account. In a correct calculation that includes the physics described by the dynamic Hubbard model   (b), the top of the band shifts upward by the relaxation energy Eq. (60b)  and the band narrows as given by 
Eq. (62). }
\label{figure1}
\end{figure} 

However, the DFT calculation does not take into account the local relaxation of the orbital when an electron is removed from this band,  since on the average all orbitals remain doubly occupied for
 infinitesimal hole doping. This has been pointed out particularly by Fulde and 
coworkers \cite{fulde,fulde2,fulde3} as a fundamental limitation of band structure calculations within the local density approximation in 
 density functional theory. Instead, a correct calculation that includes the orbital relaxation physics described by the dynamic Hubbard model Eq. (53) or its effective low energy Hamiltonian
 Eq. (57) would yield as final energy
\bmath
\beq
E_{final}^{DH}=E_{final}^{DFT}-\epsilon_R
\eeq
with
\beq
\epsilon_R=\bar{E}(1)-E(1)
\eeq
\emath the orbital relaxation energy discussed in the previous sections. Hence 
\beq
\Delta^{DH}=\Delta^{DFT}-\epsilon_R
\eeq
may become negative as shown on the right panel in Fig. 7, indicating that doped holes $will$ go into this band, contrary to the predictions of density functional theory.

In addition to the shift in the position of the energy band, density functional calculations also will miss the fact that the orbital relaxation will narrow the band \cite{dynh4}
due to the reduced overlap matrix element between expanded and unexpanded orbital. When the band is almost full the hopping amplitude from Eq. (57b) is
$t_h=S^2t$ and hence the bandwidth is
\beq
D_h=S^2D
\eeq
as shown schematically in Fig. 7. The fact that band structure calculations miss this band narrowing effect has also been emphasized recently by Casula and coworkers\cite{casula}.

For hydrogen-like atoms within the Hartree approximation the orbital relaxation energy is given by Eq. (12a) which yields $\epsilon_R=1.33eV$. Note however that
we can also obtain the orbital relaxation energy from
\beq
\epsilon_R=\frac{1}{2}(U_{bare}-U_{eff})
\eeq
from Eqs. (23) and (13). For real hydrogen-like ions we can obtain $U_{eff}$ from the difference of appropriate ionization energies and obtain\cite{dynh1}
\beq
U_{bare}-U_{eff}=4.15 eV
\eeq
within $10\%$ for $Z$ between $1$ and $8$ ($U_{bare}-U_{eff}=4.15eV, 4.10eV, 4.22eV, 4.22eV, 4.21eV, 4.19eV$, $4.15eV$, $4.05eV$ for $Z=1, 2, ...8$).
Therefore,
\beq
\epsilon_R\sim 2.1eV
\eeq
for hydrogen-like ions, about $60\%$ higher than in the Hartree approximation.

For negatively charged oxygen ions as in the $Cu-O$ planes of high $T_c$ superconductors the relaxation energy is likely to be even larger.
We can estimate \cite{tang2}  the bare $U$ from the Slater integral \cite{slater} $F^o(2p,2p)\sim 20.5eV$ for $O^o$ (the spherically averaged Coulomb repulsion between
two electrons in the $p$ shell) and the effective $U$ from the electron affinities of $O_o$ and $O^-$
\bmath
\beq
E(O^-)-E(O)=-1.45eV
\eeq
\beq
E(O^{2-})-E(O^-)=8.75eV
\eeq
\emath
yielding
\beq
U_{eff}=E(O^{2-})+E(O)-2E(O^-)=10.2eV
\eeq
and hence for the relaxation energy
\beq
\epsilon_R=\frac{1}{2}(U_{bare}-U_{eff})=5.1 eV
\eeq

Band structure calculations using density functional theory predict that in high $T_c$ cuprates the band resulting from overlap of oxygen
$p_\pi$ orbitals in the plane oriented perpendicular to the $Cu-O$ bond is several $eV$ below the Fermi energy\cite{bs1}. Instead, we have argued that to
explain high temperature superconductivity in the cuprates it is necessary that doped holes occupy the oxygen $p_\pi$ band rather than the
$Cu-Op_\sigma$ band usually assumed \cite{holesc,tang}. From the arguments presented here we conclude that the orbital relaxation energy will shift the 
$Op_\pi$ band several $eV$ upward, hence it is plausible to conclude that doped holes will occupy that band, contrary to the
band structure calculation predictions that don't take orbital relaxation energy into account \cite{goddard}.

\section{summary and discussion}

We have argued that the conventional Hubbard model that ignores the fact that atomic orbitals expand with increasing electron occupation misses
important physics of real materials, that a simple extension of the Hubbard model that we call ``dynamic Hubbard model''  takes into account.
We have analyzed in detail a dynamic Hubbard model with a single atomic orbital and an auxiliary spin $1/2$ degree of freedom. Other possible realizations of
dynamic Hubbard models are a purely electronic model with two orbitals per site and a single orbital model with an auxiliary harmonic oscillator degree of freedom \cite{holeelec}.
The effects described by dynamic Hubbard  models will be quantitatively more important for bands arising from conduction through
negatively charged ions.

We considered  the energies of the various states involved in hydrogen-like atoms, and how to model these atoms with various alternative forms of a  dynamic Hubbard Hamiltonian with 
an auxiliary spin $1/2$ degree of freedom. The effective Hamiltonian resulting from the different versions of the model is the same as far as the low energy degrees of freedom are concerned, 
with different mappings of the Hamiltonian parameters to the atomic parameters.

Finally, we have argued that conventional band structure calculations miss two important properties of electronic energy bands for cases where the bands are full or almost full
that are described by dynamic Hubbard models: 
(1) the position of the band is higher and (2) the width of the band is smaller than predicted by band structure calculations.

Thus, we argue  that  to learn about the properties of the simplest possible solid composed of hydrogen-like atoms, when there is more than one electron per atom
(band more than half full),  it is inappropriate both to use the conventional Hubbard model, conventional band structure calculations, or band structure
calculations combined with dynamical mean field theory of the conventional Hubbard model\cite{kotliar}. Instead one should 
solve the dynamic Hubbard model Hamiltonian Eq. (53) with one of the versions of the site Hamiltonians $H_i$ discussed, for parameters in the model
obtained by the mapping to the atomic parameters discussed in Sect. III, or alternatively the low energy effective Hamiltonian Eq. (57).
Exact diagonalization methods \cite{dynh6}, quantum Monte Carlo \cite{bou} or dynamical mean field theory \cite{dynhdmft} can be used to study these Hamiltonians. The physics that results, which in particular
includes superconductivity when the band is almost full \cite{holesc} and Eq. (58) is 
satisfied, is very different than the physics obtained by the conventional methods and we believe it reflects the physics
of the real system, while the physics obtained from the conventional methods does not. Of course the same considerations apply to more complicated real materials.

 We have proposed 
that this model could apply to the high $T_c$ cuprates provided doped holes go into a full oxygen $p_\pi$ band\cite{tang,holesc}. Conventional band structure calculations place this band several
$eV$ below the Fermi energy \cite{bs1}, however, we have shown here that the orbital relaxation effect described by dynamic Hubbard models will shift the position of full bands
or nearly full bands upward by
several $eV$. Hence   this supports the possibility that doped holes in the high $T_c$ cuprates may occupy the $O{p _\pi}$ orbitals rather than the 
$O{p_\sigma}$ orbitals as usually assumed. A detailed analysis of this model for 
the high $T_c$ cuprates including the $Cu-Op_\sigma$ band is given in \cite{cuprates}.

Superconductivity through this mechanism is favored both by having a low $U_{eff}$ and
 by having a small $S$. Both $U_{eff}$ and $S$ become smaller when the 
 effective ionic charge $Z$ becomes smaller (Eqs. (22) and (24)). While we have
 shown this explicitly here only for hydrogen-like ions, it is clear that it will also 
 be the case in general. Thus, superconductivity is favored by having negatively
 charged ions. The fact that negative ions have smaller on-site Coulomb repulsion
 should also be favorable for superconductivity within other superconductivity mechanisms
 that rely on electron pairing. For example, within the conventional BCS-electron-phonon interaction mechanism $\mu ^*$ should be smaller
 for systems with negative ions, leading to a higher $T_c$. It is surprising that this has not been pointed
 out before in the literature to our knowledge. This suggests for example that
 hole-doped $LiBC$ should have a substantially $lower$ transition temperature
 than $MgB_2$ because in substituting $C$ for $B$ and $Li$ for $Mg$, half of the ions in the planes 
 where conduction occurs become $C^o$ rather than $B^-$, thus
 increasing $Z$ from $1$ to $2$ and as a consequence increasing the effective Coulomb repulsion on the $C$ sites.
 Instead,  it has been predicted, based on electron-phonon calculations that presumably don't take this
 effect into account,  that
 hole-doped $LiBC$ should have a substantially $higher$ transition temperature
 than $MgB_2$ \cite{libc}.


\begin{references}
\bibitem{mott}  M. Imada, A. Fujimori and Y. Tokura,   Rev. Mod. Phys. {\bf 70}, 1039 (1998).
\bibitem{slater} J.C. Slater. ``Quantum Theory of Atomic Structure'', Mc Graw Hill, New York, 1960.

\bibitem{har1} D.R. Hartree, 
``The Wave Mechanics of an Atom with a Non-Coulomb Central Field. Part I. Theory and Methods'', 
Math. Proc. of the Cambridge Philos. Soc. {\bf 24}, 89-110, 111-132, 426-437 (1928).
\bibitem{har2} D.R. Hartree, ``The Calculation of Atomic Structures'', John Wiley $\&$ Sons, New York, 1957.
\bibitem{inapp} J. E. Hirsch,   Physica B {\bf 199$\&$200}, 366 (1994).

 \bibitem{holeelec} J. E. Hirsch, 
Phys. Rev. B {\bf 65}, 184502 (2002).


\bibitem{helium} J. Hutchinson, M. Baker and F. Marsiglio, Europ. J. of Phys. {\bf 34}, 111 (2013).

\bibitem{dynh1}  J.E. Hirsch,  Phys.Rev. Lett.  {\bf 87}, 206402 (2001).
\bibitem{dynh2}  J.E. Hirsch, Phys. Lett. A {\bf 134}, 451 (1989). 


 \bibitem{dynh3}   J.E. Hirsch,   Physica C  {\bf 201}: 347-361 (1992).
 \bibitem{dynh4}  J.E. Hirsch, Phys. Rev. B {\bf 43}, 11400 (1991).
 \bibitem{dynh5}   J.E. Hirsch, Phys. Rev. B {\bf 62} 14487  and 14998 (2000).
 \bibitem{dynh6}   J.E. Hirsch, Phys. Rev. B {\bf 65}, 214510 and  {\bf 66}, 064507 (2002).
 

\bibitem{dynh7} J.E. Hirsch, Rev. B {\bf 67}, 035103 (2003).

\bibitem{dynh8} J.E. Hirsch,  Physica C {\bf 364-365}, 37 (2001).


\bibitem{holesc} J.E. Hirsch and F.  Marsiglio,   Phys. Rev. B {\bf 39}, 11515 (1989). 

\bibitem{kinenergy} J.E. Hirsch and F.  Marsiglio, Phys. Rev. B {\bf 62},15131 (2000).





\bibitem{exp1} J.E. Hirsch, Phys.Rev. B {\bf 87} 184506 (2013).
\bibitem{exp2} J.E. Hirsch,
Physica Scripta {\bf 88}, 035704 (2013). 

\bibitem{polaronic} J.E. Hirsch, Phys. Rev. {\bf 47}, 5351 (1993).

\bibitem{incoh0} H. Ding et al, Phys. Rev. Lett. {\bf 87}, 227001 (2001).
\bibitem{incoh1} K.M. Shen et al, Phys. Rev. Letters {\bf 93}, 267002 (2004).
\bibitem{incoh2} N. Trivedi, Nature Physics {\bf 4}, 163 (2008).

\bibitem{fulde} P. Fulde,   Int. J. Quantum Chem. {\bf 76}, 385 (2000).
\bibitem{fulde2} A. Stoyanova, L. Hozoi, P. Fulde and H. Stoll,   J. Chem. Phys. {\bf 131}, 044119 (2009).

\bibitem{fulde3} 
L. Hozoi, U. Birkenheuer, P. Fulde, A. Mitrushchenkov, and H. Stoll , Phys. Rev. B {\bf 76}, 085109 (2007).

\bibitem{casula} M. Casula, Ph. Werner, L. Vaugier, F. Aryasetiawan, T. Miyake, A. J. Millis, and S. Biermann, Phys. Rev. Lett. {\bf 109}, 126408 (2012).
\bibitem{tang2} J.E. Hirsch and S. Tang, Phys.Rev. B {\bf 40}, 2179 (1989).

\bibitem{bs1} W.E. Pickett,  Rev. Mod. Phys. {\bf 61}, 433 (1989).

\bibitem{tang} J.E. Hirsch and S. Tang, Sol.St. Comm. 69: 987(1989). 
\bibitem{goddard} 
This possibility was also proposed early on by Goddard and coworkers: 
Y. Guo, J.M. Langlois and W.A.  Goddard,  Science {\bf 239}, 896 (1988).

\bibitem{kotliar} G. Kotliar et al,   Rev. Mod. Phys. {\bf 78}. 865 (2006).

\bibitem{bou} K. Bouadim, M. Enjalran, F. HŽbert, G. G. Batrouni, and R. T. Scalettar,
Phys. Rev. B{\bf 77}, 014516
(2008)
 
 \bibitem{dynhdmft}  G.H. Bach, J.E. Hirsch and  F. Marsiglio,,   Phys. Rev. B {\bf 82}, 155122 (2010). 

\bibitem{cuprates} J.E. Hirsch, arXiv:1407.0042 (2014).

\bibitem{libc} H. Rosner and W.E. Pickett, Phys. Rev. Lett. {\bf 88}, 127001 (2002).


 \end{references}
\end{document}